\documentclass[aps,prb,showpacs,twocolumn]{revtex4}
\usepackage{amssymb}
\usepackage{amsmath}
\usepackage{graphicx}
\usepackage{dcolumn}
\usepackage{bm}
\usepackage{units}
\usepackage{booktabs}
\usepackage{multirow}

\hfuzz=\maxdimen
\begin{document}

\title{Thermoelectric effects in FeCo$|$MgO$|$FeCo magnetic tunnel junctions}
\author{Shizhuo Wang}
\affiliation{Department of Physics, Beijing Normal University, Beijing 100875, China}
\author{Ke Xia}
\affiliation{Department of Physics, Beijing Normal University, Beijing 100875, China}
\author{Gerrit E. W. Bauer}
\affiliation{Institute for Materials Research and WPI-AIMR, Tohoku University, Sendai
980-8577, Japan\\
and Kavli Institute of NanoScience, Delft University of Technology, 2628 CJ
Delft, The Netherlands}
\date{\today }

\begin{abstract}
We studied the thermoelectric coefficients (Seebeck and thermal conductance)
of FeCo$|$MgO$|$FeCo(001) magnetic tunnel junctions (MTJs) from first
principles using a generalized Landauer-B\"{u}ttiker formalism. FeCo$|$MgO$|$%
FeCo(001) MTJs usually yield smaller thermoelectric effects compared with
epitaxial Fe$|$MgO$|$Fe(001) MTJs. The (magneto-) Seebeck effect is
sensitive to the details of the FeCo$|$MgO interfaces. Interfacial oxygen
vacancies (OVs) can enhance the thermoelectric effects in MTJs greatly. We
also compute angular dependent Seebeck coefficients that provide more
information about the transport physics. We report large deviations from the
Wiedemann-Franz law at room temperature.
\end{abstract}

\maketitle

\section{Introduction}

Spin caloritronics is a research direction~\cite{STB2012,Gerrit2012} that
provides alternative strategies for thermoelectric waste heat recovery and
cooling. Seebeck~\cite{KU2008} and Peltier~\cite{JF2012} effects in magnetic
nanostructures become spin-dependent, \textit{i.e.}, different spin channels
contribute differently and can be modulated by the magnetization direction.
Moreover, in magnetic heterostructures a thermal spin transfer torque (TST)~%
\cite{hatami2007,jxtL2011} can be induced by heat currents.

Magneto thermoelectric effects in magnetic tunnel junctions (MTJs) were
measured recently~\cite{walter2011,AB2013,lie2011,Hu2012} partly motivated
by its potential applications in magnetic random access memory devices. The
Seebeck rectification effect in MTJs might be beneficial for scavenging
waste heat.~\cite{Hu2012} The reported Seebeck coefficients ($S$) in MgO
based MTJs vary from 22 $\mu $V/K (Ref.~\onlinecite{Hu2012}) to -770 $\mu $%
V/K (Ref.~\onlinecite{AB2013}) for similar barrier thicknesses while the
difference in Seebeck coefficients between magnetic parallel and
antiparallel configurations, $\Delta S$=$S_{p}-S_{ap}$ were measured from
-8.7$\mu $V/K (Ref.~\onlinecite{walter2011}) to -272$\mu $V/K (Ref.~%
\onlinecite{lie2011}). Seebeck coefficients as high as mV/K, have also been
reported.~\cite{Lin2012,JM2013}

Due to the difficulty in determining the temperature difference across the
tunneling barrier, the intrinsic Seebeck coefficient cannot be measured
directly but has to be determined via thermal modelling, which introduces
uncertainties. The calculations based on realistic electronic band
structures yield Seebeck coefficients less than 60$\mu $V/K at room
temperature (RT).~\cite{MC2012,CH2013}

For MTJs, the energy dependence of the conductance is sensitive to the band
alignment between insulator and metal. Small changes in the computational
procedures and parameters can result in quite different thermoelectric
coefficients. In this paper, we address the complications in order to
increase the accuracy of the predictions and find out how large the Seebeck
coefficients might become as well as its tunability by interface engineering.

The Landauer-B\"{u}ttiker formalism has been generalized to thermal
transport and to thermoelectric cross-effects by Butcher,~\cite{butcher1990}
which treats electrical transport in terms of transmission through a
scattering region between electron reservoirs. Knowing the energy dependent
conductance, the Seebeck coefficient and electric thermal conductance can be
calculated.

In this paper, we combine the Landauer-B\"{u}ttiker formalism for spin
polarized thermal and electrical transport with realistic electronic band
structures to compute the Seebeck coefficient and thermal conductance in
FeCo-MgO MTJs. In Sec. II we present the details of the formalism. In Sec.
III the method is used to calculate the thermoelectric coefficients of FeCo$%
| $MgO$|$FeCo with perfect interfaces and in the presence of oxygen
vacancies (OVs). In Sec. IV we summarize our results.

\section{Thermoelectric Coefficients}

We model a device sandwiched by left (L) and right (R) electron leads with
chemical potential difference $\Delta \mu =\mu _{L}-\mu _{R}$ and
temperature bias $\Delta T=T_{L}-T_{R}$. The heat flow $\overset{.}{\emph{Q}}
$ and electric current \emph{I} then read\cite{Groot}
\begin{equation}
\binom{\Delta \mu /e}{\dot{Q}}=\left(
\begin{array}{cc}
R & S \\
\Pi & -\kappa%
\end{array}%
\right) \binom{I}{\Delta T},
\end{equation}%
where $R$ is the electrical resistance and the Seebeck coefficient $S$ and
Peltier cooling coefficient $\Pi $ are related by the Onsager-Kelvin
relation $\Pi =ST$\textit{.}

The spin-dependent conductance

\begin{equation}
G_{\sigma }=\frac{e^{2}}{h}\int d\epsilon t_{\sigma }\left( \epsilon \right) %
\left[ -\partial _{\epsilon }f(\epsilon )\right] ,  \label{G}
\end{equation}%
where $\sigma =\uparrow (\downarrow )$ denotes spin, $t_{\sigma }\left(
\epsilon \right) $ is the energy-dependent spin transmission probability,
the Fermi occupation $f(\epsilon ,\mu ,T)=\left[ e^{\left( \epsilon -\mu
\right) /k_{B}T}+1\right] ^{-1}$ is a function of energy $\epsilon $,
electrochemical potential $\mu =(\mu _{L}+\mu _{R})/2$, and temperature $%
T=(T_{L}+T_{R})/2$, and we here define $f(\epsilon )=f(\epsilon ,\mu ,T)$.

In the linear response approximation, total electric current reads
\begin{widetext}
\begin{eqnarray}
I\left( \Delta T,\Delta V\right) &=& \frac{e}{h}\sum_{\sigma }\int
d\epsilon \left[
f\left( \epsilon ,\mu _{L},T_{L}\right) -f\left( \epsilon
,\mu_{R},T_{R}\right)
\right] t_{\sigma }\left( \epsilon \right)  \notag \\
&=& -\frac{e\Delta T}{hT}\sum_{\sigma }\int d\epsilon \left(
\epsilon -\mu \right) \partial _{\epsilon }f(\epsilon )t_{\sigma
}\left( \epsilon \right)-\frac{e^{2}}{h}\Delta V\sum_{\sigma }\int
d\epsilon \partial _{\epsilon }f(\epsilon )t_{\sigma }\left(
\epsilon \right) .  \label{Icurrent}
\end{eqnarray}
\end{widetext}

The Seebeck coefficient $S=-\left( \Delta V/\Delta T\right)
_{I=0}$ is obtained by setting $I=0$ in Eq. (\ref{Icurrent}) as

\begin{eqnarray}
S &=&\frac{1}{eT}\frac{\sum_{\sigma }\int d\epsilon \left[ \left( \epsilon
-\mu \right) t(\epsilon )\partial _{\epsilon }f(\epsilon )\right] }{%
\sum_{\sigma }\int d\epsilon t(\epsilon )\partial _{\epsilon }f(\epsilon )}
\notag \\
&=&\frac{S_{\uparrow }G_{\uparrow }+S_{\downarrow }G_{\downarrow }}{%
G_{\uparrow }+G_{\downarrow }},  \label{S}
\end{eqnarray}%
where
\begin{equation}
S_{\sigma }=\frac{1}{eT}\frac{\int d\epsilon \left[ \left( \epsilon -\mu
\right) t_{\sigma }(\epsilon )\partial _{\epsilon }f(\epsilon )\right] }{%
\int d\epsilon t_{\sigma }(\epsilon )\partial _{\epsilon }f(\epsilon )},
\end{equation}%
and $e=-1.602\times 10^{-19}$C. When the energy dependent conductance varies
slowly around fermi level, one can use the Sommerfield expansion and Eq. (%
\ref{S}) becomes

\begin{equation}
S=eL_{0}T\left[ \partial _{\epsilon }\ln G(\epsilon )\right] |_{E_{f}},
\label{Szhankai}
\end{equation}%
with Lorenz number $L_{0}\equiv \pi ^{2}\left( k_{B}/e\right)
^{2}/3=2.45\times 10^{-8}$V$^{2}$K$^{-2}$.

There are two parallel channels of heat transport, viz. the
electrons and the phonons. We calculate only the electron part in
the presence of a nonequilibrium thermal distribution. The
electronic contribution to the thermal conductance $\kappa $ is
defined as

\begin{equation}
\kappa \equiv -\left( \frac{\dot{Q}}{\Delta T}\right) _{I=0}=-\left(
K+S^{2}GT\right) .  \label{kappa}
\end{equation}
where $K$ is given in the Landauer-B\"{u}ttiker formalism by Ref.~\onlinecite%
{butcher1990} as

\begin{eqnarray}
K &=&\frac{{k_{B}}^{2}T}{e^{2}}\int d\epsilon G\left( \epsilon \right)
\left( \frac{\epsilon -\mu }{k_{B}T}\right) ^{2}\partial _{\epsilon
}f(\epsilon ).
\end{eqnarray}
At low temperatures, the leading term in the Sommerfield expansion $K$ is

\begin{equation}
K=-L_{0}TG\left( \epsilon _{f}\right) ,  \label{K}
\end{equation}%
We may disregard the term $S^{2}GT$ when $S^{2}\ll L_{0}$, which leads to
the Wiedemann-Franz (WF) relation

\begin{eqnarray}
\kappa &\approx &L_{0}TG\left( \epsilon _{f}\right).  \label{WF}
\end{eqnarray}

The tunnel magneto resistance ratio (TMR) is defined in terms of the
conductances for parallel (P) and anti-parallel (AP) configurations:

\begin{equation}
\text{TMR\textbf{=}}\frac{\text{G}_{\text{P}}-\text{G}_{\text{AP}}}{\text{G}%
_{\text{AP}}}\times \text{100\%,}
\end{equation}
where $G_{\text{P/AP}}=\frac{e^{2}}{h}\sum_{\sigma }t_{\text{P/AP}}^{\sigma
}(\epsilon _{f})$.

Similarly tunnel magneto-Seebeck (TMS) and tunnel magneto heat resistance
ratios (TMHR) as

\begin{equation}
S_{\text{m}}\text{\textbf{=}}\frac{S_{\text{P}}-S_{\text{AP}}}{\min \left(
\left\vert \text{S}_{\text{P}}\right\vert \text{,}\left\vert \text{S}_{\text{%
AP}}\right\vert \right) }\times \text{100\%},  \label{Sm}
\end{equation}%
and

\begin{equation}
\kappa _{\text{m}}\text{\textbf{=}}\frac{\kappa _{\text{P}}-\kappa _{\text{AP%
}}}{\min \left( \kappa _{\text{P}},\kappa _{\text{AP}}\right) }\times \text{%
100\%}.  \label{KM}
\end{equation}

At sufficiently low temperature, the WF relation may be used in Eq. (\ref{KM}%
) and the value of $\kappa _{m}=$ TMR.

\section{THERMOELECTRICS OF $\mbox{FeCo$|$MgO$|$FeCo}$(001)}

\subsection{Model}

\begin{figure}[b]
\includegraphics[width=8.5cm]{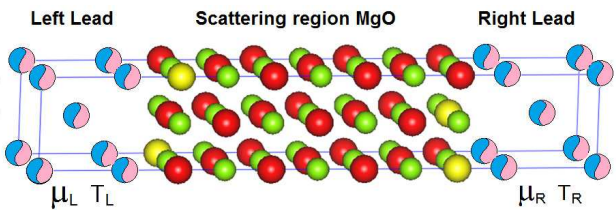}
\caption{(Color online) Schematical atomic structure of the FeCo$|$MgO$|$%
FeCo magnetic tunnel junction. The two FeCo$|$MgO interfaces contain
randomly distributed oxygen vacancies (Roughness would be generated by
fluctuating thickness of the barrier). FeCo random alloy: blue-magenta
spheres; Mg: green spheres; O: red spheres. The O vacancies places at yellow
sphere positions.}
\label{fig1}
\end{figure}

We consider a two-probe device consisting of a MgO \newline
barrier and two semi-infinite ferromagnetic leads as shown in Fig.
\ref{fig1}. The electric current is applied along the (001) growth
direction, The atoms at interfaces are kept unrelaxed in their
bulk bcc positions. Oxygen vacancies (OVs) in MgO are
energetically favorable because they relax the compressive strain
at the interface during crystal growth.\cite{PG} We assume that
OVs only exists at the first atomic layer at the interface between
the MgO barrier and FeCo.

The thermoelectric coefficients are governed by the energy dependence of the
conductance. While the lateral supercell method can be used to handle the
impurity scattering in metallic system,~\cite{Ke06} the required high
accuracy of the energy-dependent conductance would be difficult to obtain
for the present system, since in MTJs, the error bar due to the disorder
configurations usually is much larger than that in metallic systems because
of the absence of self-averaging over the Brillouin Zone (BZ).

The density functional theory calculation with coherent potential
approximation (CPA) is therefore more suitable for a quantitative
theoretic analysis of spin transport through
Fe$_{x}$Co$_{1-x}|$MgO$|$Fe$_{y}$Co$_{1-y} $ MTJs, where $x, y$
are numbers between 0 and 1. The transport properties are
evaluated here by the Keldysh nonequilibrium Green function
including nonequilibrium vertex corrections.\cite{ke08} The method
is generalized to handle non-collinear magnetization similar to
non-collinear problem in scattering wave function
method.\cite{wang2008}

We use 4$\times$10$^4$ k points in the full two-dimensional BZ to
ensure excellent numerical convergence. Other details of the
electronic structure and transport calculation can be found in
Ref.~\onlinecite{ke08}. Our CPA method can only handle disorder in
the scattering region; we use virtual crystal approximation (VCA)
to deal with the potential functions in the alloy leads. To prove
VCA is a qualified method, we study the
Fe$|$FeCo$|$MgO$|$FeCo$|$Fe MTJs with alloy FeCo in the scattering
region, which 6 monolayers(6ML) of FeCo is enough to add coherent
potential as leads, then we compare the energy-dependent
conductance between FeCo$|$MgO(6ML)$|$FeCo (VCA) with
Fe$|$FeCo(6ML)$|$MgO(6ML)$|$FeCo(6ML)$|$Fe. Our calculations shows
that results are not sensitive to this simplification of the
electronic structure of the leads.

\begin{table}[tbh]
\caption{TMR ratio of FeCo$|$MgO($n$ML)$|$FeCo with clean and dirty (5\%OV
at both) interfaces for 3, 5, 7 and 9 monolayers, respectively.}
\label{tab:tmrcal}
\begin{center}
\begin{tabular*}{8cm}{@{\extracolsep{\fill}}cccc}
\hline\hline
$n$ & concentration & disorder & TMR(\%) \\ \hline
3 & Fe$_{0.25}$Co$_{0.75}$ & clean & 577 \\
& Fe$_{0.50}$Co$_{0.50}$ & clean & 934 \\
& Fe$_{0.75}$Co$_{0.25}$ & clean & 1003 \\
& Fe$_{0.50}$Co$_{0.50}$ & 5\%OVs & 209 \\ \hline
5 & Fe$_{0.25}$Co$_{0.75}$ & clean & 853 \\
& Fe$_{0.50}$Co$_{0.50}$ & clean & 900 \\
& Fe$_{0.75}$Co$_{0.25}$ & clean & 1017 \\
& Fe$_{0.50}$Co$_{0.50}$ & 5\%OVs & 113 \\ \hline
7 & Fe$_{0.25}$Co$_{0.75}$ & clean & 902 \\
& Fe$_{0.50}$Co$_{0.50}$ & clean & 957 \\
& Fe$_{0.75}$Co$_{0.25}$ & clean & 1061 \\
& Fe$_{0.80}$Co$_{0.20}$ & clean & 1178 \\
& Fe$_{0.50}$Co$_{0.50}$ & 5\%OVs & 90 \\
Exp.~\cite{lie2011} & Co$_{0.6}$Fe$_{0.2}$B$_{0.2}$ &  & 70$\thicksim $%
140(RT) \\
Exp.~\cite{SI2008} & Co$_{0.2}$Fe$_{0.6}$B$_{0.2}$ &  & 604(RT), 1144(5K) \\
\hline
9 & Fe$_{0.25}$Co$_{0.75}$ & clean & 947 \\
& Fe$_{0.50}$Co$_{0.50}$ & clean & 1033 \\
& Fe$_{0.75}$Co$_{0.25}$ & clean & 1101 \\
& Fe$_{0.50}$Co$_{0.50}$ & 5\%OVs & 82 \\ \hline
Exp.~\cite{walter2011} & Fe$_{0.50}$Co$_{0.50}$ &  & 330(RT) \\ \hline\hline
\end{tabular*}%
\end{center}
\end{table}

The TMR ratios calculated for different barriers in our calculation are
compared with experiments in Table \ref{tab:tmrcal}. 3ML is the thinnest MgO
barrier achievable by current experiment technique.\cite{JC2013} In
experiments,\cite{SI2008} TMR ratios can be maximized through controlled
annealing and other grow conditions reaching our theoretical values for the
clean interfaces. However, most likely the samples used in the
thermoelectric experiments\cite{walter2011,lie2011} with lower TMR ratios
contains 3\%OVs$\sim$5\%OVs.

\begin{figure}[b]
\includegraphics[width=8.5cm]{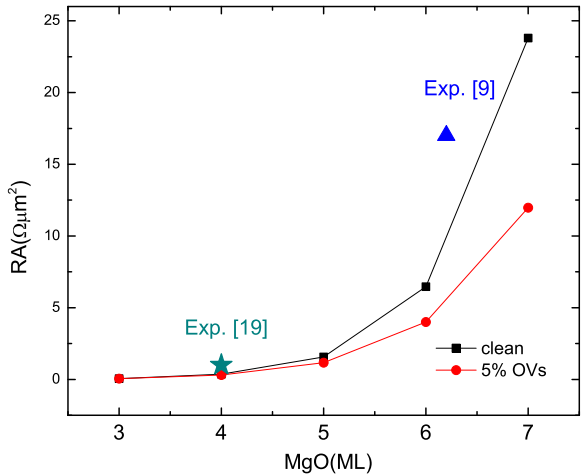}
\caption{ (Color online) Resistance-area (RA) dependence on the thickness of
MgO barriers with clean (black-square) and disorder (with 5\%OVs)
(red-circle) Fe$_{0.5}$Co$_{0.5}|$MgO interfaces. Blue-up and cyan-star are
the experiment values.}
\label{RA}
\end{figure}

In Fig. \ref{RA}, we compare RA with published experiments. For a 7ML thick
MgO barriers (1.6nm) with PC, our calculation yields 23.8$\Omega\mu$m$^{2}$%
(clean) and 12$\Omega\mu$m$^{2}$(5\%OVs), close to the measured junction
resistance of 17 $\Omega\mu$m$^{2}$ for 1.5nm thick tunnel junctions.\cite%
{lie2011}

\subsection{Energy dependent conductance $G_{\protect\sigma }\left( \protect%
\epsilon \right)$}

\begin{figure}[b]
\includegraphics[width=8.5cm]{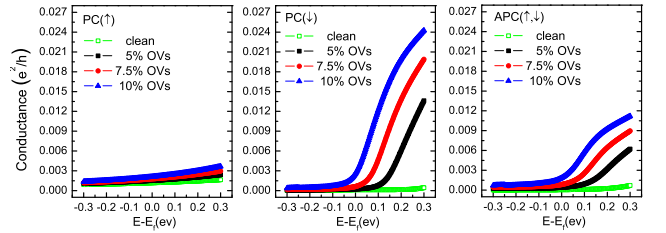}
\caption{ (Color online) Energy-dependent conductance of Fe$_{0.5}$Co$%
_{0.5}| $MgO(5ML)$|$Fe$_{0.5}$Co$_{0.5}$ with clean (green hollow squares),
5\%OVs (black squares), 7.5\%OVs (red circles), 10\%OVs (blue up-triangles)
at both interfaces with P and AP configurations, respectively.}
\label{5fecopc}
\end{figure}

Even though both Fe$|$MgO$|$Fe and FeCo$|$MgO$|$FeCo MTJs show large TMR
ratios, their spectral conductance is quite different. Resonant transmission
channels exist just below the Fermi level in Fe-MgO based MTJs,\cite%
{jxtL2011} but not in FeCo-MgO based MTJs. Fig. \ref{5fecopc} shows the
energy dependence of the conductance of Fe$_{0.5}$Co$_{0.5}|$MgO(5ML)$|$Fe$%
_{0.5}$Co$_{0.5}$ (001) MTJs with given concentration of OVs at both
interfaces for both PC and APC. The energy window in the plots corresponds
to 11k$_B$T at room temperature (300K), where k$_B$ is the Boltzmann
constant. The slope of the energy-dependent transmission around the Fermi
energy dramatically changes by only small amounts of OVs. The APC shows a
similar tendency as PC shown in the Fig. \ref{5fecopc}.

\begin{figure}[t]
\includegraphics[width=8.5cm]{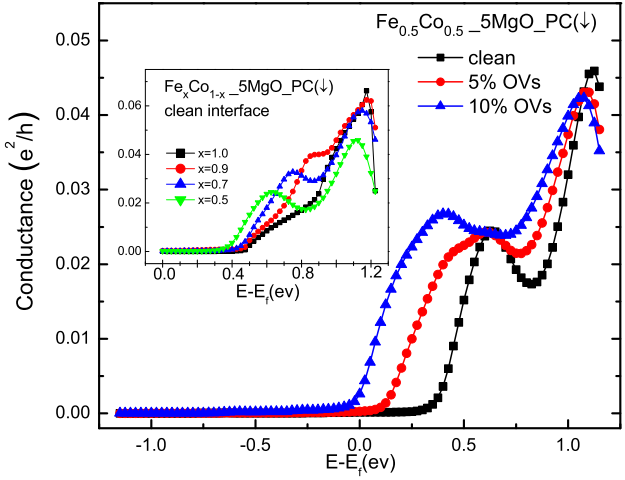}
\caption{(Color online) Energy-dependent conductance for minority-spin
channel in Fe$_{0.5}$Co$_{0.5}|$MgO(5ML)$|$Fe$_{0.5}$Co$_{0.5}$ MTJs with
clean (black squares), 5\%OVs (red circles), 10\%OVs (blue up-triangles)
interface for P configurations, respectively. Inset: energy-dependent
conductance for minority-spin channel in Fe$_{x}$Co$_{1-x}|$MgO(5ML)$|$Fe$%
_{x}$Co$_{1-x}$ MTJs with $x=1$ (black squares), 0.9 (red circles), 0.7
(blue up-triangles), 0.5 (green down-triangles), respectively.}
\label{5down1ev}
\end{figure}

In order to understand this, we show results for a wider energy window of $E=E_{f}\pm $%
1.2eV in Fig. \ref{5down1ev}. Two peaks exist above the Fermi level for
minority-spin in Fe$_{0.5}$Co$_{0.5}|$MgO(5ml)$|$Fe$_{0.5}$Co$_{0.5}$(001)
MTJs. The OVs broaden the peaks and shift their towards the Fermi level up
to a certain amount also enhance the conductance around the Fermi level.

Since thermoelectric effects are closely related to the slope of
energy-dependent conductance near the Fermi level, a proper amount of OVs at
FeCo$|$MgO interfaces enhance the Seebeck and Peltier constants.

The origin of these two peaks for MTJs with random alloys is not obvious.
Fortunately, similar conductance peaks also exist in epitaxial Fe$|$MgO$|$Fe
MTJs. The inset of Fig. \ref{5down1ev} shows the energy-dependent
conductance of MTJs with different alloy concentrations (including pure Fe)
as electrodes. For epitaxial Fe$|$MgO$|$Fe (black square), there is one
clear peak exist at 1eV above the Fermi energy and a shoulder around 0.8eV.
with 10\% Co atoms doping (red circles), the shoulder develops into a
plateau. At higher Co concentrations (x=0.7, 0.5), the plateau becomes a
second peak near the Fermi level.

\begin{figure}[t]
\includegraphics[width=8.5cm]{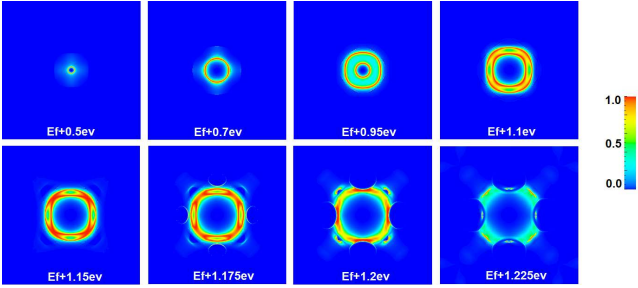}
\caption{ (Color online) k$_{\Vert }$ resolved transmission for
minority-spins in Fe$|$MgO(5ML)$|$Fe MTJs in the parallel configurations and
(001) direction with clean interfaces for different energies.}
\label{Reson}
\end{figure}

\begin{figure}[tbp]
\includegraphics[width=8.5cm]{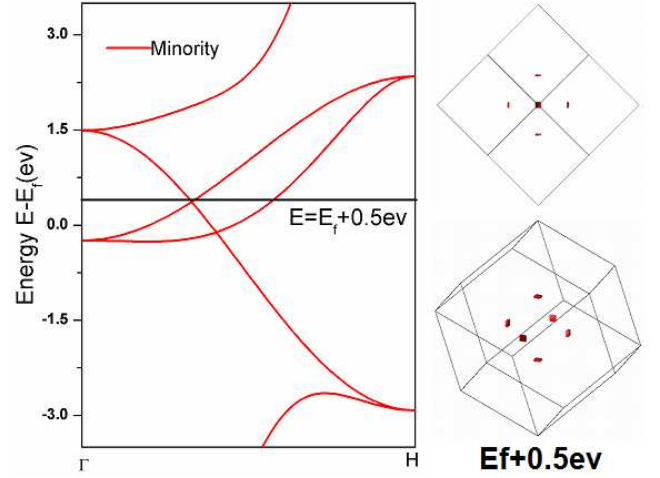}
\caption{ (Color online) Minority-spin band structure of Fe and its energy
surface at E$_{f}$+0.5eV in the reduced BZ, respectively.}
\label{feband}
\end{figure}

For epitaxial Fe$|$MgO$|$Fe MTJs, we can identify the origin of these
conductance peaks by inspecting the band structure of Fe. We plot the k$%
_{\Vert }$ resolved transmission for the minority-spin bands in Fe$|$MgO(5ML)%
$|$Fe MTJs at different energies in Fig. \ref{Reson}. From $E\sim E_{f}+$%
0.5eV to $E\sim E_{f}+$1.2eV, we observe a "hot" ring with energy-dependent
diameter. The maximum transmission can reach unity, which is evidence for
resonant tunneling channels that emerge form six small symmetry "bubbles" in
constant energy surface at $E\sim E_{f}+$0.5eV as shown in Fig. \ref{feband}%
. These bubbles have sharp edges that result in two resonant hot concentric
rings. As these "bubbles" get larger with energy, the diameter of the hot
rings increases. At even higher energies, these "bubbles" hybridize with
other transmission channels and loose their resonant character.

\subsection{Seebeck coefficient $S$}

\begin{figure}[t]
\includegraphics[width=8.5cm]{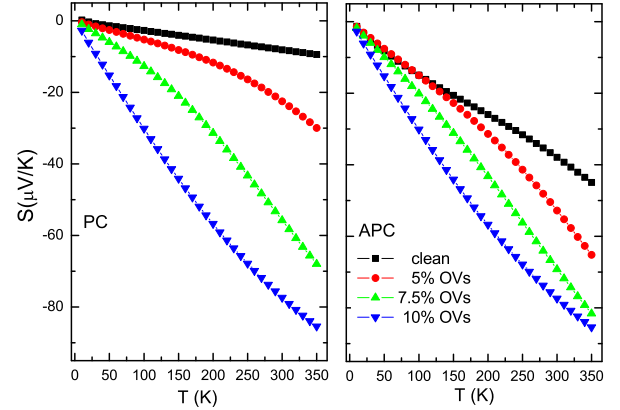}
\caption{(Color online) Seebeck coefficients of Fe$_{0.5}$Co$_{0.5}|$MgO(5ML)%
$|$Fe$_{0.5}$Co$_{0.5}$ MTJs with clean interfaces (black squares), 5\% (red
circles), 7.5\%(green up-triangles), 10\%OVs (blue down-triangles) at both
interfaces.}
\label{5fecos}
\end{figure}

Seebeck coefficients can be obtained by Eq. (\ref{S}), the energy range of
integration depending on the temperature. We choose $E=E_{f}\pm $0.3eV in
this paper, which corresponds to a precision of more than 99.5\% for T=300K.
Fig. \ref{5fecos} exhibits the Seebeck coefficient in Fe$_{0.5}$Co$_{0.5}|$%
MgO(5ML)$|$Fe$_{0.5}$Co$_{0.5}$ MTJs with different concentration of OVs at
the FeCo$|$MgO interface. The clean interface (black squares) in MTJs gives
the smallest Seebeck coefficient, while an increasing OVs enhances the
effect until 10\%OVs (blue down-triangles) by an order of magnitude for PC.
The magneto-Seebeck $S_{\text{m}}$ is 369.3\% and -3.6\% for clean and
10\%OVs at both interfaces, respectively. Moreover, we give Seebeck
coefficients and the corresponding magneto-Seebeck ratio for different MgO
barriers in Tab. \ref{tab:seecal}.

Firstly, the Seebeck coefficient gets larger with thicker MgO barriers with
identical interface disorder for both PC and APC. For example, the Seebeck
coefficient of MTJs with 9 monolayers MgO is 2$\sim$10 times larger than
that of with 3 monolayers for different interfacial quality and
configurations, whereas the conductance changes by 5 orders of magnitude.
Seebeck coefficients are clearly not sensitive to MgO barrier thickness.

Secondly, the sign of the Seebeck coefficient does not change with thickness
at RT, and the value of Seebeck coefficient is enhanced by the OVs at the
interface for a certain layers of MgO. In our study, the thermoelectrical
effect are maximized for 10\%OVs for 5 monolayers MgO barrier, which can be
understood from Fig. \ref{5down1ev}.

Thirdly, the order and sign of the magneto-Seebeck ratio ($S{_{\text{m}}}$)
is sensitive to the details of the interfacial roughness. Take the
calculated results for 5 monolayers MgO in Tab. II and Fig. \ref{5fecopc} as
an example. When the interface is clean, the APC has larger Seebeck
coefficient than the PC, while the magneto-Seebeck is large and positive.
When the interface contains some OVs, both PC and APC display a larger
thermoelectric effect. But $S_{\text{P}}$ always grows faster than $S_{\text{%
AP}}$, which can be seen by inspecting the slopes around the Fermi level in
Fig. \ref{5fecopc}. So $S{_{\text{m}}}$ changes sign in some cases lead to a
very small $S{_{\text{m}}}$. For samples with a small $S{_{\text{m}}}$ at
RT, we expect the sign will change at low temperature.

The existence of OVs at the interfaces has a great effect on the slope of
the energy-dependent conductance at the Fermi level, which means we can tune
thermoelectric effects by the concentration of OVs at the interfaces. Our
calculation of Seebeck coefficients and TMS of 9ML MgO barriers give an
estimation with 7\%OVs at the interfaces are consistent with the experiment
results~\cite{walter2011} in Tab. \ref{tab:seecal} for PC and APC,
respectively.

\begin{table}[tbp]
\caption{\ Seebeck coefficients (in unit of $\protect\mu$V/K) and
magneto-Seebeck S$_{m}$(\%) of Fe$_{0.5}$Co$_{0.5}|$MgO($n$ML)$|$Fe$_{0.5}$Co%
$_{0.5}$ MTJs at T=300K for P and AP, and we compared them with experiments
results in Ref. \onlinecite{walter2011}, respectively.}
\label{tab:seecal}
\begin{center}
\begin{tabular*}{8cm}{@{\extracolsep{\fill}}ccccc}
\hline\hline
$n$ & disorder & P & AP & $S_{\text{m}}$(\%) \\ \hline
3 & clean & -2.09 & -23.82 & 1039.7 \\
& 5\%OVs & -6.93 & -19.41 & 180.1 \\
& 7.5\%OVs & -12.87 & -16.65 & 29.4 \\
& 10\%OVs & -14.78 & -25.77 & 74.4 \\ \hline
5 & clean & -8.08 & -37.92 & 369.3 \\
& 5\%OVs & -22.48 & -52.79 & 134.8 \\
& 7.5\%OVs & -55.80 & -69.23 & 24.1 \\
& 10\%OVs & -77.36 & -74.70 & -3.6 \\ \hline
7 & clean & -15.13 & -50.26 & 267.6 \\
& 5\%OVs & -40.46 & -76.44 & 88.9 \\
& 7.5\%OVs & -101.33 & -99.20 & -2.2 \\
& 10\%OVs & -124.93 & -98.93 & -26.3 \\ \hline
9 & clean & -23.12 & -61.50 & 166.0 \\
& 5\%OVs & -62.79 & -99.80 & 58.9 \\
& 6.5\%OVs & -112.86 & -119.05 & 8.4 \\
& 7\%OVs & -132.10 & -124.15 & -6.4 \\
& 7.5\%OVs & -149.17 & -127.99 & -16.5 \\
& 10\%OVs & -155.79 & -121.74 & -30.0 \\ \hline\hline
Exp.~\cite{walter2011} &  & -107.9 & -99.2 & -8.8 \\ \hline\hline
\end{tabular*}%
\end{center}
\end{table}

Although MTJs with 9 monolayers MgO have the largest Seebeck coefficient as
shown in Table \ref{tab:seecal}. However, since its conductance and
thermoelectric current $GS\Delta T$ is so small, which would result in lower
thermoelectric current, and 3MgO junctions still generate the largest
thermoelectric power for a given temperature difference.

\begin{figure}[t]
\includegraphics[width=8.5cm]{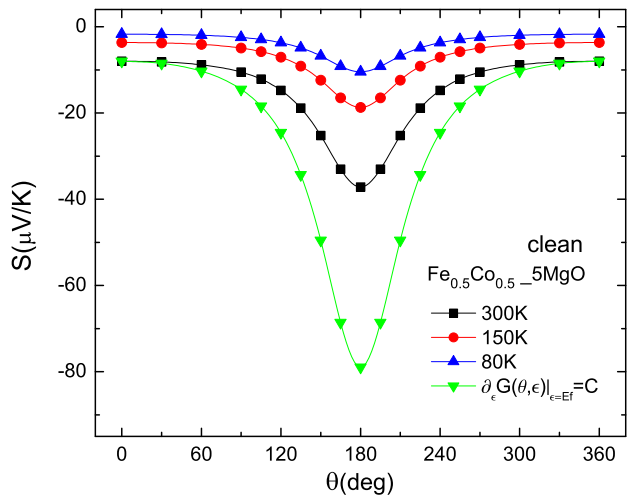}
\caption{ (Color online) Angle dependent Seebeck coefficient of Fe$_{0.5}$Co$%
_{0.5}|$MgO$|$Fe$_{0.5}$Co$_{0.5}$ under environment temperatures 300K
(black quare), 150K (red circle), 80K (blue up-triangle), function-fitting
for 300K (green down-triangle) with clean interface, respectively.}
\label{5sangle}
\end{figure}

The angular dependent Seebeck coefficient (ADSC) and conductance can provide
more information about the transport process. We compute the ADSC at 300K
(black squares), 150K (red circles), 80K (blue up-triangles) for Fe$_{0.5}$Co%
$_{0.5}|$MgO(5ML)$|$Fe$_{0.5}$Co$_{0.5}$ in Fig. \ref{5sangle}. The
horizontal axis denotes the relative angle of magnetization between the two
leads. The Seebeck coefficient varies slowly from PC to 90$^{\circ }$, and
drastic from 90$^{\circ }$ to APC. The ADSC looks is consistent with the
report.\cite{AGS2013}

This unusual variation deviates strongly from the familiar trigonometric
dependence for thick layers. The phenomenon can be simply explained as
follows, the angle dependent conductance could be read as $G\left( \theta
\right) =\frac{1}{2}\left( G_{pc}+G_{apc}\right) +\frac{1}{2}\left(
G_{pc}-G_{apc}\right) \cos \theta $, and Eq. (\ref{Szhankai}) as $S\left(
\theta \right) =eL_{0}T\frac{\partial _{\epsilon }G\left( \theta ,\epsilon
\right) }{G\left( \theta ,\epsilon \right) }|_{\epsilon =E_{f}}$. If $%
\partial _{\epsilon }G\left( \theta ,\epsilon \right) |_{\epsilon =E_{f}}$
is constant \newline
for different angle, $S\left( \theta \right) $ behave as 1/cos$\theta $. If,
on the other hand $\partial _{\epsilon }G\left( \theta ,\epsilon \right)
|_{\epsilon =E_{f}}\sim $cos$\theta $, the Seebeck coefficient becomes
constant as $S_{\text{P}}$. We compare a Seebeck coefficients at 300K (black
squares) with 1/$\cos \theta $ (green down-triangles) in Fig. \ref{5sangle}.
The similar behavior can be interpreted as evidence for AGSC.

\subsection{Thermal conductance $\protect\kappa $}

\begin{figure}[tbp]
\includegraphics[width=8.5cm]{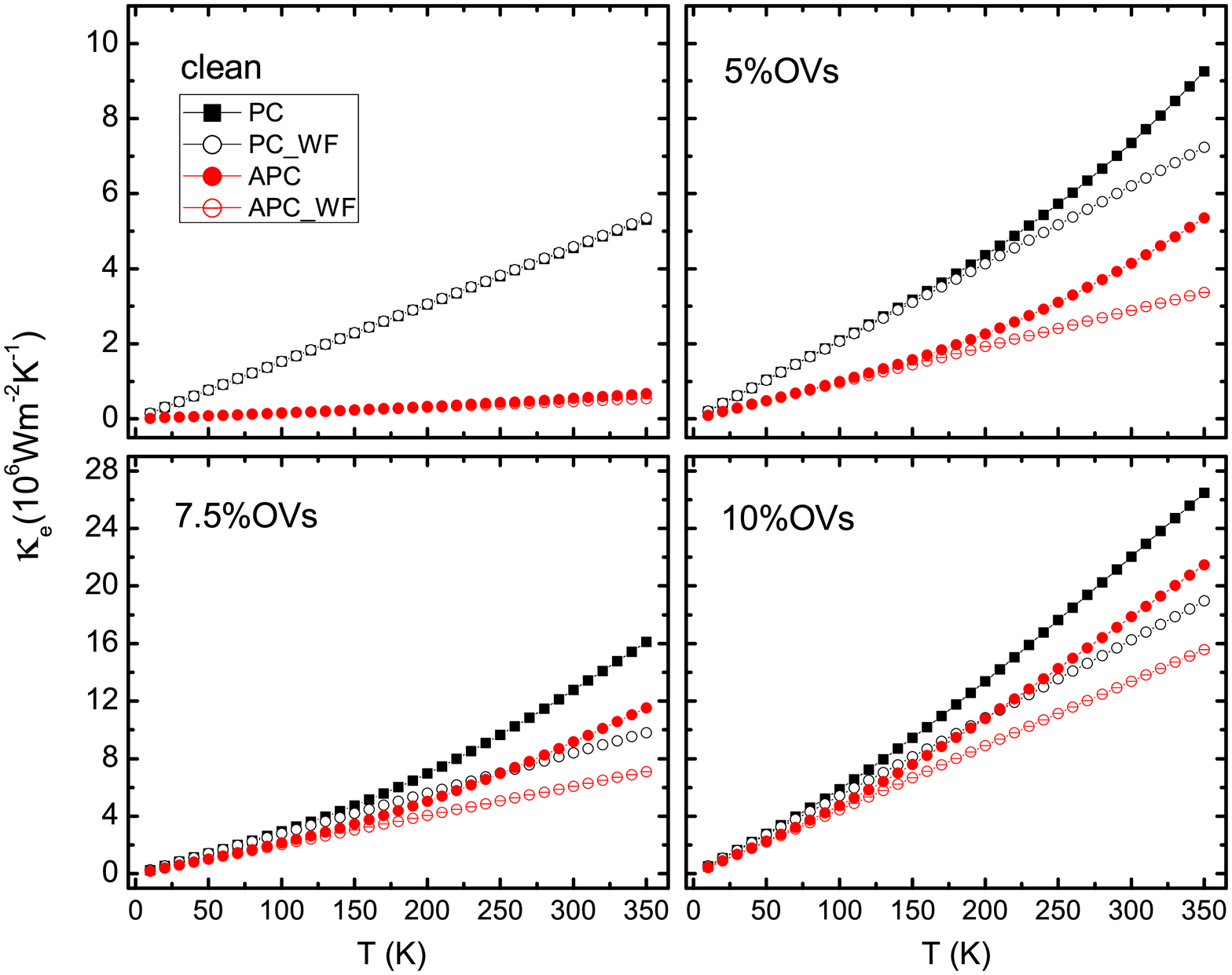}
\caption{ (Color online) Temperature dependent thermal conductance for Fe$%
_{0.5}$Co$_{0.5}|$MgO(5ML)$|$Fe$_{0.5}$Co$_{0.5}$ clean interface and 5\%,
7.5\%, 10\%OVs at both interface,the hollow black and hollow red line
denotes P and AP configurations, respectively.}
\label{5fecok}
\end{figure}

The electronic heat conductance depends on the symmetric component of the
spectral around the Fermi level. Fig. \ref{5fecok} shows the thermal
conductance of Fe$_{0.5}$Co$_{0.5}|$MgO(5ML)$| $Fe$_{0.5}$Co$_{0.5}$(001)
with 0\% 5\%, 7.5\%, 10\%OVs at both interfaces, respectively. The thermal
conductance is increased by the OVs similar to the charge conductance and
Seebeck coefficient. 10\% OVs enlarge the thermal conductance by 5 and 33
times for P and AP configurations compared to clean interfaces at RT,
respectively. The tunnel magneto heat resistance ratios (TMHR) is 744.4\%
and 23.3\% for clean and 10\%OVs at both interfaces, respectively. The order
of TMHR changes greatly by the OVs at the interface.

\begin{figure}[t]
\includegraphics[width=8.5cm]{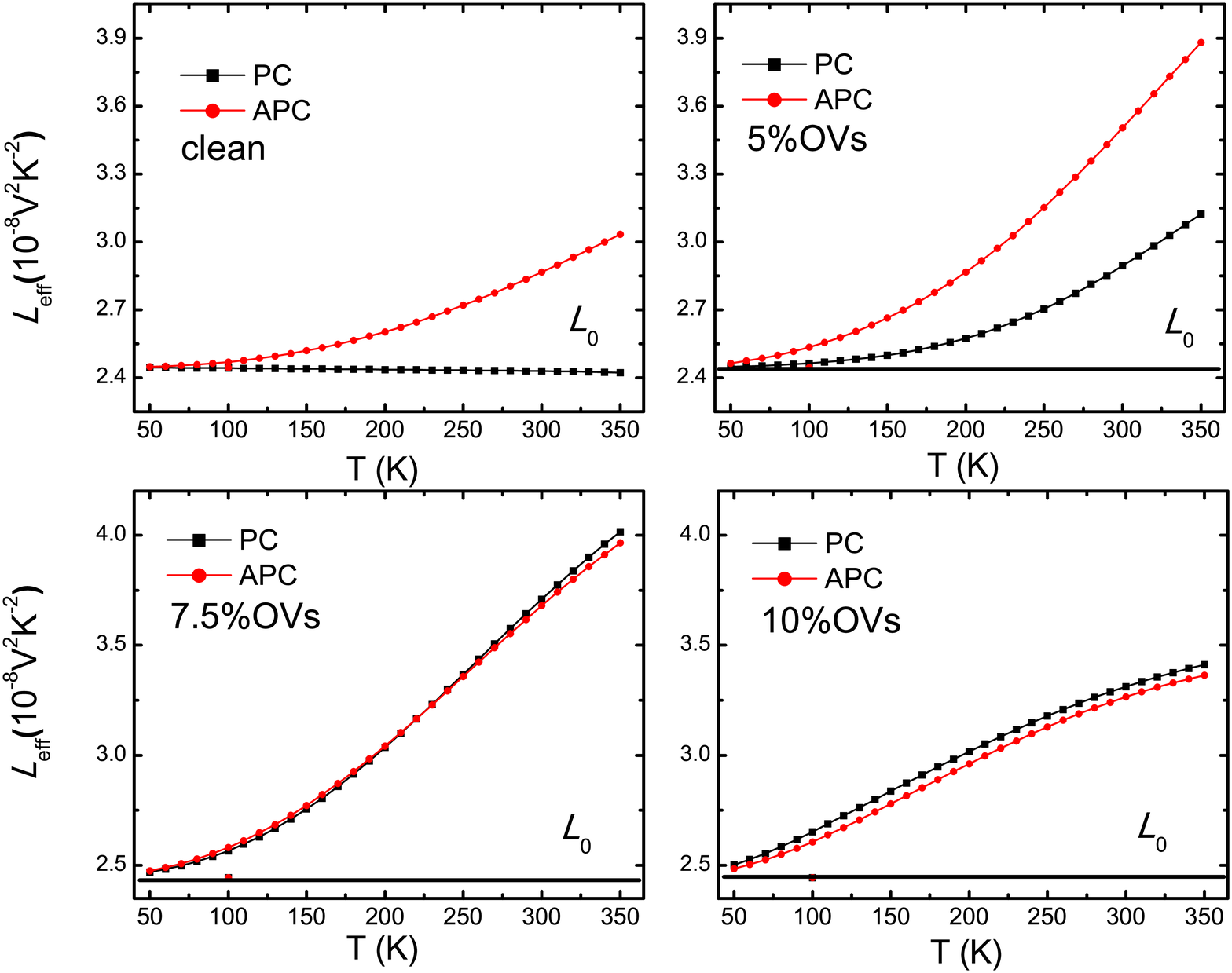}
\caption{ (Color online) Effective Lorenz number $L_{\text{eff}}$ of Fe$%
_{0.5}$Co$_{0.5}|$MgO(5ML)$|$Fe$_{0.5}$Co$_{0.5}$(001) with clean interface
and 5\%, 7.5\%, 10\%OVs at both interfaces,the black and red dot-line denotes P
and AP configurations, respectively.}
\label{5L}
\end{figure}

Additional, we tested the WF law as a function of temperature. At low
temperature, $S^{2}\ll L_{0}$ and the WF law holds. When the thermal
conductance does not vary linear with temperature due to a breakdown of the
Sommerfield approximation or the Seebeck coefficient is getting large (see
Fig. \ref{5fecos}), the WF relation is no longer valid, and deviation are
observed in in Fig. \ref{5fecok}. We define an effective Lorenz number $L_{%
\text{eff}}$ from
\begin{equation}
\kappa =L_{\text{eff}}TG\left( \epsilon _{f}\right) ,
\end{equation}%
to parameterize the study the breakdown of the WF Law by compare it with the
Lorenz constant $L_{0}=2.45\times 10^{-8}$V$^{2}$K$^{-2}$. We display the
temperature dependent effective Lorenz number for different OVs
concentrations in Fig. \ref{5L}, in which $L_{\text{eff}}$ is found to
become significantly enhanced from L$_{0}$ with increasing temperature. We
show the thermal conductance and corresponding TMHR of MTJs for different
MgO barriers in Tab. III, Firstly, the thermal conductances decrease sharply
with thicker MgO barriers and interfacial roughness for both PC and APC.
Secondly, the thermal conductance is enhanced by the interfacial OVs with a
fixed thickness, whereas, the order of tunnel magneto heat resistance ratios
is decreased. Thirdly, the order of TMHR does not change to much with the
same interfacial roughness for thicker MgO barriers.

\begin{table}[t]
\caption{Thermal conductance (in units of 10$^{6}$Wm$^{-2}$K$^{-1}$) and $%
\protect\kappa_{m}$(\%) of Fe$_{0.5}$Co$_{0.5}|$MgO($n$ML)$|$Fe$_{0.5}$Co$%
_{0.5}$ MTJs at T=300K under P and AP with different interfacial roughness,
respectively.}
\label{tab:kap}
\begin{center}
\begin{tabular*}{8cm}{@{\extracolsep{\fill}}ccccc}
\hline\hline
$n$ & disorder & P & AP & $\kappa_{m}$ \\ \hline
3 & clean & 128.67 & 16.44 & 682.7 \\
& 5\%OVs & 147.26 & 55.28 & 166.4 \\
& 7.5\%OVs & 169.91 & 84.09 & 102.1 \\
& 10\%OVs & 198.13 & 122.99 & 61.1 \\ \hline
5 & clean & 4.56 & 0.54 & 744.4 \\
& 5\%OVs & 7.35 & 4.15 & 77.1 \\
& 7.5\%OVs & 12.77 & 9.17 & 39.3 \\
& 10\%OVs & 22.03 & 17.86 & 23.3 \\ \hline
7 & clean & 0.31 & 0.037 & 737.8 \\
& 5\%OVs & 0.84 & 0.53 & 58.5 \\
& 7.5\%OVs & 2.27 & 1.59 & 42.8 \\
& 10\%OVs & 5.73 & 4.15 & 38.1 \\ \hline
9 & clean & 0.027 & 0.003 & 800.0 \\
& 5\%OVs & 0.137 & 0.087 & 57.5 \\
& 7.5\%OVs & 0.581 & 0.383 & 51.7 \\
& 10\%OVs & 2.364 & 1.552 & 52.3 \\ \hline\hline
\end{tabular*}%
\end{center}
\end{table}

\section{Summary}

In conclusion, we computed the thermoelectric coefficients of FeCo$|$MgO$|$%
FeCo MTJs from first-principles. OVs at FeCo$|$MgO interfaces can be used to
engineer thermoelectric effects. While interface disorder can greatly
increase the Seebeck coefficient, it suppresses the magneto-Seebeck ratio.
The vacancy concentration is therefore an important design parameter in
switchable thermoelectric devices based on magnetic tunnel junctions.

\section*{ACKNOWLEDGMENTS}

We gratefully acknowledges financial support from National Basic Research
Program of China (2011CB921803, 2012CB921304), NSF-China (11174037,
61376105), JSPS Grants-in-Aid for Scientific Research (25247056, 25220910),
the EU-RTN Spinicur, and DFG Priority Programme 1538 "Spin-Caloric
Transport" (GO 944/4).

\end{document}